%% file: CP060.tex
\documentclass[runningheads]{llncs}
\usepackage[]{graphics}
\usepackage[linesnumbered,ruled,vlined]{algorithm2e}
\usepackage{mathtools}
\usepackage{multirow}
\usepackage{amsmath}
\usepackage{bm}
\usepackage{subcaption}
\usepackage{tikz}
\usepackage{listings}
\usepackage{commath}
\usepackage{soul}
\usepackage[normalem]{ulem}
\usepackage{todonotes}

\usepackage{booktabs}
\usepackage{relsize}
\usepackage{multirow}
\usepackage{array}
\definecolor{tableheadcolor}{rgb}{0.8,0.8,1.0}
\definecolor{tablealtcolor}{rgb}{0.9,0.9,0.95}

\newcolumntype{B}{>{\global\let\currentrowstyle\relax}}
\newcolumntype{^}{>{\currentrowstyle}}

\newcolumntype{C}{>{\bfseries}}

\let\oldtabular\tabular
\let\endoldtabular\endtabular
\renewenvironment{tabular}{\sffamily\oldtabular}{\endoldtabular}

\usepackage{caption}
\usepackage[binary-units=true,detect-family]{siunitx}

\begin{document}
\title{Distributed Work Stealing in a Task-Based Dataflow Runtime}
%
%
\author{Joseph John\inst{1}\orcidID{0000-0002-0031-4793} \and
Josh Milthorpe\inst{1,2}\orcidID{0000-0002-3588-9896} \and
Peter Strazdins\inst{1}\orcidID{0000-0001-8541-1551}}
\authorrunning{J. John et al.}
%
\institute{Australian National University, Canberra, Australia 
\and Oak Ridge National Laboratory, Oak Ridge, TN \\
\email{\{joseph.john, josh.milthorpe, peter.strazdins\}@anu.edu.au}}
\maketitle              
\begin{abstract}
\input{paper/abstract}
\end{abstract}

\input{paper/intro}

\input{paper/contributions}
\input{paper/literature}

\input{paper/stealing_method}

\input{paper/experiments}

\input{paper/conclusion}

\input{paper/acknowledgement}

\input{bibliography}

\end{document}

%% file: paper/abstract.tex
The task-based dataflow programming model has emerged as an alternative to the process-centric programming model for extreme-scale applications.
However, load balancing is still a challenge in task-based dataflow runtimes.
In this paper, we present extensions to the PaRSEC runtime to demonstrate that distributed work stealing is an effective load-balancing method for task-based dataflow runtimes.
In contrast to shared-memory work stealing, we find that each process should consider future tasks and the expected waiting time for execution when determining whether to steal.
We demonstrate the effectiveness of the proposed work-stealing policies for a sparse Cholesky factorization, which shows a speedup of up to 35\% compared to a static division of work. 

%% file: paper/intro.tex
\section{Introduction}
\label{section: intro}




The task-based dataflow programming model has emerged as an alternative to the process-centric model of computation in distributed memory. In this model, an application is a collection of tasks with dependencies derived from the data flow among the tasks. Tasks can be executed in any order that maintains the dependency relations between
them. When compared to a process-centric model, the task-based dataflow programming model has shown more scalability as it exposes more asynchronicity within the application~\cite{Bosilca2012,Cao2020}. Also, the programmer has a global view of tasks and data, while low-level problems such as scheduling and data transfer are taken care of by the runtime. 



At present, most implementations of the task-based dataflow programming model are limited to a static work division between nodes.
This paper addresses this limitation by exploring whether distributed work stealing can be used as an automatic load balancing method in a task-based dataflow runtime. We use Parallel Runtime Scheduling and Execution Controller (PaRSEC) \cite{Bosilca2012} as the base framework. PaRSEC is a task-based dataflow runtime for distributed heterogeneous architectures. To the best of our knowledge, this is the first work in a task-based dataflow runtime in distributed memory to use distributed work stealing as a load balancing technique. 




%% file: paper/contributions.tex
\subsection{Contributions}
\label{section: contributions}

The contributions of this paper are as follows:
\begin{enumerate}
    \item We add distributed work stealing to PaRSEC runtime for automatic load balancing. 
    \item We extend the Template Task Graph (TTG) to allow the programmer to decide if a particular task can be stolen. 
    \item We introduce new victim policies based on waiting time and show that this is more efficient than the existing victim policies. 
    \item We introduce a new thief policy based on future tasks and show that this is more efficient than the existing thief policies. 
\end{enumerate}

%% file: paper/literature.tex
\section{Related Work}
\label{section: literature}


Work sharing and work stealing are two primary approaches to load balancing in task-based programming models. In work sharing, an overloaded compute node shares its work with the underloaded nodes, while in work stealing, an underloaded node steals work from the overloaded nodes. Work sharing requires information collection about the load in a set of nodes and coordination between the nodes in this set to balance the load between them. The main disadvantages of work sharing are that collecting load information may pose scalability issues, and due to the asynchronous nature of task execution there is no guarantee that the information received reflects the actual load status. On the other hand, in work stealing, a thief node initiates a steal request based on its load and the victim node chooses whether to allow the steal based on its load. Both victim and thief make independent decisions without any coordination between them.
While load-balancing in task-based runtimes was first introduced in shared memory through work stealing in Cilk~\cite{BLUMOFE199655,Friggo98}, shared memory load-balancing is not discussed here as we are only interested in load-balancing across nodes in partitioned global address space (PGAS) and distributed memory.

 
The PGAS model presents a unified global memory, logically partitioned among different nodes.  This global address space makes it possible to use global data structures, shared between nodes, to implement load-balancing strategies. In Habanero-UPC++ \cite{Kumar2017a}, each node publishes the current count of stealable tasks in a shared variable in global address space and the work stealing decisions are made based on this.
In X10, each node maintains a shared queue to hold stealable tasks and a local queue to hold non-stealable tasks~\cite{Agarwal2007,Guo2009,Paudel2013,Tardieu2012}. A starving node can directly steal from the shared queue of another node. X10 also enforces work sharing if work stealing fails \cite{Paudel2015b}. Chapel~\cite{Chamberlain2007} allows dynamic task mapping i.e. a task can be mapped to any node in the system but once the tasks are mapped to a node they cannot be stolen.



In the distributed-memory model, each node is a separate memory and execution domain.
Unlike PGAS models, there are no shared global data structures that can be leveraged for cooperation between the different nodes.
Perarnau et al.~\cite{Perarnau2014} study work stealing performance in MPI, but here the work stealing is a property not of the runtime but of the benchmark itself.
In Chameleon~\cite{KLINKENBERG202020,Klinkenberg2020}, work sharing is possible but it can happen only at global MPI synchronization points.
Samfass et al.~\cite{Samfass2018} implement work sharing in partial differential equation workloads but the work sharing is possible only between time steps.
In CnC~\cite{Schlimbach6498557} and Legion~\cite{Bauer2012}, similar to Chapel, dynamic task mapping is possible, but once mapped to a node the tasks cannot be stolen. CnC also uses a broadcast operation to locate data items and this operation is not scalable either.


Task-based dataflow programming model is a subset of a task-based programming model where the execution progression is controlled by the flow of data from one task to the next. Charm++ is a task-based dataflow runtime that supports work sharing \cite{acun2014} and it is especially well suited for iterative applications. At present, there is no dataflow task-based programming model that offers work stealing in distributed memory.

%% file: paper/stealing_method.tex
\section{Adding Work Stealing to PaRSEC}
\label{section: stealing_method}

PaRSEC\footnote{https://josephjohn@bitbucket.org/josephjohn/parsec\_ttg.git} is a task-based dataflow runtime, where the execution of tasks is fully distributed, with no centralized components. Each task in PaRSEC is an instance of a task class and all tasks that belong to a particular task class have the same properties except the data it operates on and its unique id. PaRSEC supports multiple domain-specific languages (DSL) and these DSLs help the user define the different task classes in a program, as well the dependency relations between the tasks.  In this paper, we focus on the Templated Task Graph (TTG) DSL~\cite{Bosilca2020TheTT} as it can better handle irregular applications. An application can be called irregular if it has unpredictable memory access, data flow or control flow. To study whether work stealing is effective in a task-based dataflow runtime, we added an extra module \textit{migrate} to PaRSEC to do all operations related to work stealing. We also changed how tasks are described in TTG, to support work stealing.

The migrate module uses a dedicated \emph{migrate} thread for all stealing related activities.
The thread is created when the PaRSEC communication module is initialized and destroyed when the termination detection module in PaRSEC detects distributed termination.
All communication to and from the migrate module is carried out using the PaRSEC communication module.
The migrate thread constantly checks the state of the node and transitions the node to a \textit{thief} if it detects starvation. On detecting starvation, the thief node sends a steal request to a \textit{victim} node.
The victim's migrate thread processes the steal request and selects tasks to be migrated to the thief node. 
When a task is selected as a victim of a steal request, the input data of the victim task are copied to the thief node and the victim task is recreated in the thief node. To implement this functionality,
we added a new function \texttt{migrate} to the task class. The migrate thread invokes this function to copy the input data to the thief node. Once all data have arrived, the thief recreates the victim task, with the same unique id, and it is treated like any other task by the thief node.

\input{paper/task_description}


\subsubsection{Thief policy}
\input{paper/thief_policy}

\subsubsection{Victim Policy}
\input{paper/victim_policy}

%% file: paper/task_description.tex
\subsubsection{New Task Description}
\label{new_task_description}

To give the user control over which tasks can be stolen, we introduced another wrapper function in TTG\footnote{https://github.com/josephjohnjj/ttg.git} (Listing \ref{lst: ttg_wrap2}),  which takes a function \texttt{is\_stealable} as an additional argument (The details about the wrapping function are available in ~\cite{Bosilca2020TheTT}). For instance, in a sparse linear algebra computation, tasks of the same type may operate on a dense or sparse tile. So the programmer may decide that tasks that operate on a sparse tile cannot be stolen. 

\begin{lstlisting}[language = C++, caption = New TTG wrapping function, label = {lst: ttg_wrap2}]
ttg::wrapG(task_body, is_stealable, input_edges, 
  output_edges, task_name, input_edge_names, output_edge_names);
\end{lstlisting}

The function \texttt{is\_stealable}  has the same signature as the task body, and it has access to the same data as the task body. 


%% file: paper/thief_policy.tex
The thief policy dictates two aspects of stealing: 1) How is a victim node selected?
and 2) What qualifies as starvation in a node?
Perarnau et al.~\cite{Perarnau2014} demonstrated that randomised victim node selection is best suited for distributed work stealing, so we use the same in this paper. A naive approach to work stealing only consider the ready tasks waiting for a worker thread as the indicator for available load in a node and if the \textbf{available} ready task is zero, starvation is assumed. 
We show that this is not the correct way to predict starvation as stealing takes non-zero time, and in that time new tasks can be activated in a starving node. So, we propose that along with ready tasks we should also consider the tasks that will be scheduled in the near future to measure starvation. We take the successors of the tasks in execution as the future tasks. Based on these we tested two starvation policies:

\begin{enumerate}
    \item Ready tasks only: a steal request is initiated if there are no currently ready tasks.
    \item Ready tasks + Successor tasks: a steal request is initiated if there are no currently ready tasks and no local successors of tasks currently in execution.
\end{enumerate}

%% file: paper/victim_policy.tex
Victim policies impose an upper bound on the number of tasks allowed to be stolen by a thief node.  We test three victim policies:

\begin{enumerate}
    \item Half: Half the stealable tasks are allowed to be stolen per steal request.
    \item Chunk: An arbitrary number of stealable tasks is allowed to be stolen per steal request (we went with a chunk size of 20 as it is half of the total worker threads available).
    \item Single: Only one stealable task is allowed to be stolen per steal request (this is a special case \textit{chunk}, where the chunk size is 1).
\end{enumerate}

 The victim policy does not guarantee work stealing. For instance, if there are 40 stealable tasks available, the victim policy \textit{Half} requests the scheduler to return as many tasks as possible up to a maximum of 20. This is not guaranteed to yield a task, as the migrate thread competes with worker threads, and the worker threads may end up getting all the available tasks. So the victim policy makes the best effort to migrate a permissible number of stealable tasks, with an upper bound on the number of tasks migrated.
 
 At present, the waiting time of the task is not considered when permitting a steal. In this paper, the victim policies have an additional condition: work stealing is allowed only if the time required to migrate the task to the thief node is less than the time the task has to wait for a worker thread. The waiting time is calculated as follows:

\[ average\; task\; execution\; time = \frac{execution\; time \; elapsed }{tasks\; executed\; till\; now} \]
\[ waiting\; time\;  = ( \frac{\#ready\; tasks}{\#worker\; threads} \;+\; 1 ) \;*\;  average\; task\; execution\; time \]

%% file: paper/experiments.tex
\section{Experiments}
\label{section: experiments}

The experiments were conducted on the Gadi supercomputer in the National Computing Infrastructure, Australia. Each node on Gadi has two 24-core Intel Xeon Scalable Cascade Lake processors with \SI{3.2}{\giga\hertz} clock speed and \SI{192}{\gibi\byte} of memory. All the experiments were run using openmpi (v4.0.2),  intel-mkl (v2020.2.254) and intel-compiler (v2020.2.254).


As there is only one MPI process per node,  \textit{node} and \textit{process} are used interchangeably in this section. All the experiments are conducted using 40 worker threads per node. To confirm that execution times follow a normal distribution we performed D'Agostino-Pearson and Shapiro-Wilk tests. We also conducted an analysis of variance (ANOVA) test on the execution times with and without work stealing, to confirm that the two groups come from different distributions. 



\input{paper/experiment/benchmarks}

\input{paper/experiment/thief}

\input{paper/experiment/victim}



%% file: paper/experiment/benchmarks.tex
\subsection{Benchmarks}
\label{section: Benchmarks}

We use Cholesky factorization on a tiled sparse matrix as the benchmark to measure the different aspects of work stealing. In this benchmark, the matrix is divided into tiles and each tile is either sparse (filled with zeroes) or dense. In our runs, exactly half of the tiles are dense and tiles are cyclically distributed across nodes. We chose Cholesky factorization as the benchmark because it is a good representative of linear algebra benchmarks, and it has been used extensively to study various aspects of distributed computing including work-stealing. Also, there are 4 types of tasks in Cholesky factorization – POTRF, GEMM, TRSM and SYRK. The different task types have different execution times for the same tile size, presenting a challenge for distributed work-stealing.

We also used the Unbalanced Tree Search (UTS) benchmark \cite{Olivier2006UTSAU} to study the victim policies. In the UTS benchmark, different trees can be created by configuring the different features of the benchmarks. 

\subsection{Potential for Work Stealing }
\label{Effectiveness of Work Stealing}

\begin{figure}[htbp]
  \centering
  \includegraphics[width=\textwidth]{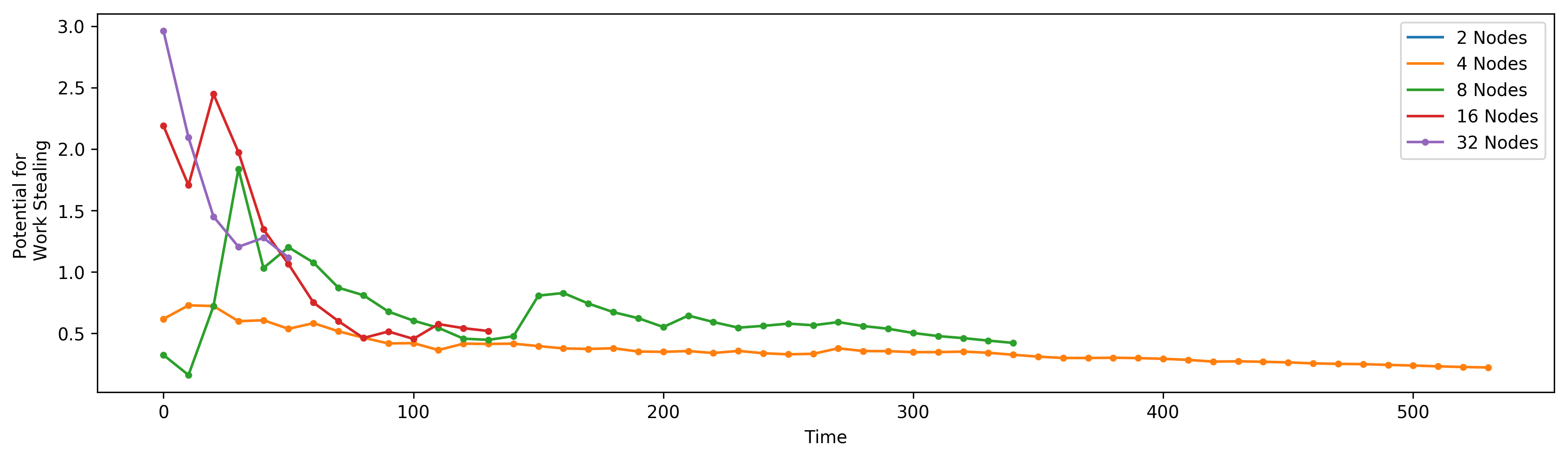}
  \caption{Potential for work stealing when using an interval size of 10 seconds. (Global matrix of $10000^2$ 64-bit elements, organized as $200^2$ tiles of $50^2$ elements)}
  \small
  \label{fig: instant_effective_all}
\end{figure}

Intuitively, task stealing is most effective when there is a workload imbalance and when there are active thief nodes. To quantify the potential for work stealing as the computation progresses, we divided the execution time of the benchmarks without work stealing into intervals of equal duration. Within each interval, whenever a worker thread successfully executed a \textit{select} operation, the number of ready tasks were polled. Using these polled ready tasks, the potential for work stealing $E^{b}$ in the interval $b$ for $P$ processes is calculated as:

\setlength{\abovedisplayskip}{3pt} \setlength{\abovedisplayshortskip}{3pt}
\setlength{\belowdisplayskip}{4pt} \setlength{\belowdisplayshortskip}{4pt}

\begin{equation} \label{eq1}
\begin{split}
E^b &= I^b  * P
\end{split}
\end{equation}

where $I^b$ is the workload imbalance in the interval $b$, calculated as:

\begin{equation} \label{eq2}
\begin{split}
I^b &=  max(w_1^b, w_2^b, ..., w_P^b) -\frac{\sum_{i=1}^{P} w_i^b}{P}\\
\end{split}
\end{equation}

where $w_i^b$ is the workload of process $i$ in the interval $b$, calculated as:

\begin{equation} \label{eq3}
\begin{split}
w_i^b &= \frac{ \frac{\sum_{j=1}^{N} o_j^b}{N_{b}} }{max(o_1^b, o_2^b, ..., o_{N_{b}}^{b})}
\end{split}
\end{equation}

where $o_j^b$ is the jth polled value in interval $b$ and $N_b$ is the total number of polled values in interval $b$. Fig.~\ref{fig: instant_effective_all} gives the potential for work stealing obtained experimentally for the different intervals for the different number of nodes. From Fig.~\ref{fig: instant_effective_all}, we see that the work stealing has the most potential at the beginning of the execution for all numbers of nodes, remaining highest for 8 nodes as the execution progresses.

%% file: paper/experiment/thief.tex
\subsection{Thief Policy}
\label{subsection: thief}

\begin{figure}[t]
  \centering
  \includegraphics[width=.5\textwidth]{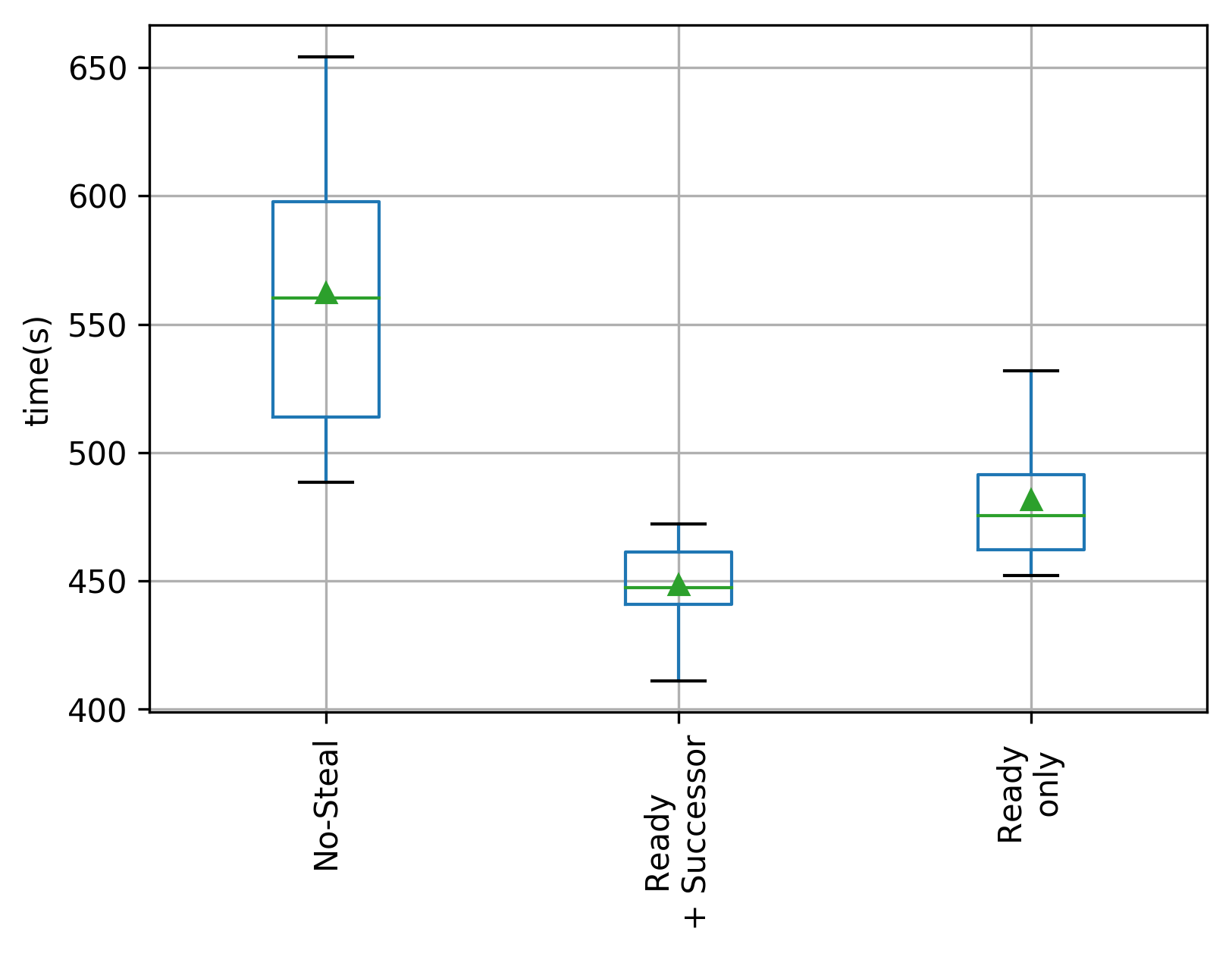}
  \caption{Thief policies that counts only ready tasks versus policy that counts ready and successor tasks. (Global matrix of $10000^2$ 64-bit elements, organized as $200^2$ tiles of $50^2$ elements. Four nodes, \textit{Single} victim policy)}
  \small
  \label{fig: successor}
\end{figure}

The experiments on thief policy show that performance of work stealing is better when future tasks are taken into consideration to determine starvation. Fig.\ref{fig: successor} shows the performance of a thief policy that uses only ready tasks to determine starvation, against a thief policy that use ready tasks as well as future tasks (`No-Steal' in the experiments refer to the experimental runs without using work stealing).  Here, the successor tasks of tasks currently in execution are taken as future tasks. From the figure, we observe that the performance of work stealing is better if future tasks are taken into consideration when determining starvation.

\begin{figure}[t]
  \centering
  \includegraphics[width=\textwidth]{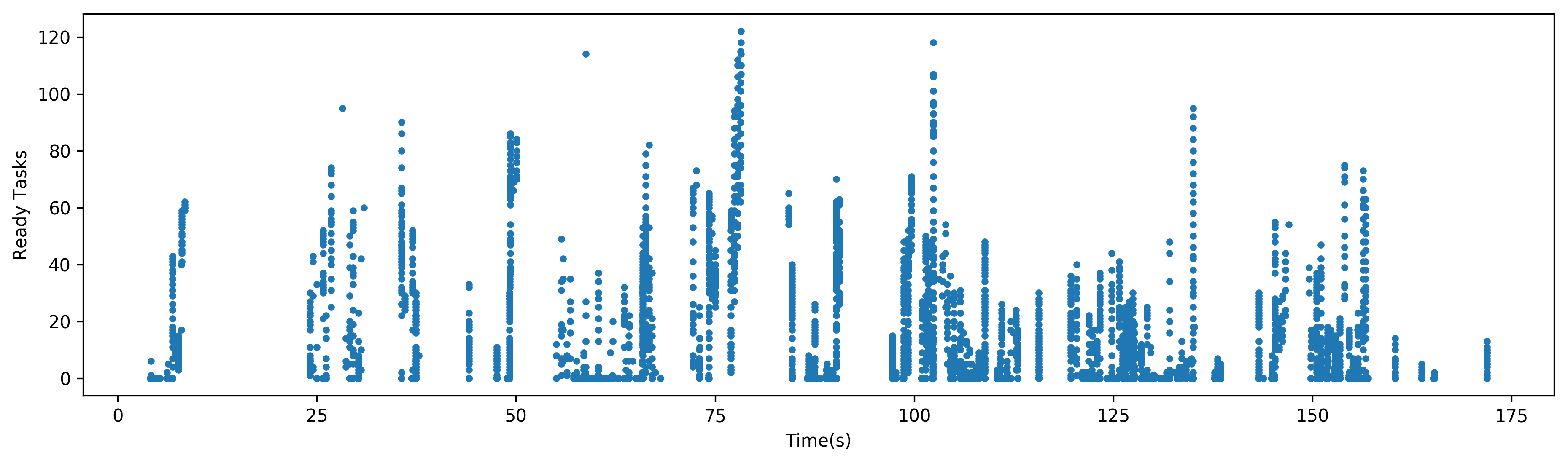}
  \caption{Ready tasks in a thief node when a stolen task arrives. Only ready tasks were considered to determine starvation. (Global matrix of $10000^2$ 64-bit elements, organized as $100^2$ tiles of $100^2$ elements; two nodes) }
  \small
  \label{fig: thief node}
\end{figure}

To understand why work stealing underperforms while using only ready tasks to determine starvation, we counted the ready tasks in a thief node when a stolen task arrives. Fig. \ref{fig: thief node} shows the result of this experiment and we can see that when the task arrives the number of ready tasks in the thief node is quite high. This  means that the stolen task will have to wait a substantial amount of time before it is selected for execution. This happens because even when there are no ready tasks in a thief node, there may still be tasks in execution, each of which can have multiple successor tasks. So by the time a stolen task arrives, the tasks in execution may have added their successors to the ready queue.

%% file: paper/experiment/victim.tex
 \subsection{Victim Policy}
 \label{subsection: victim}
 
\begin{figure}[t]
  \centering
  \includegraphics[width=\textwidth]{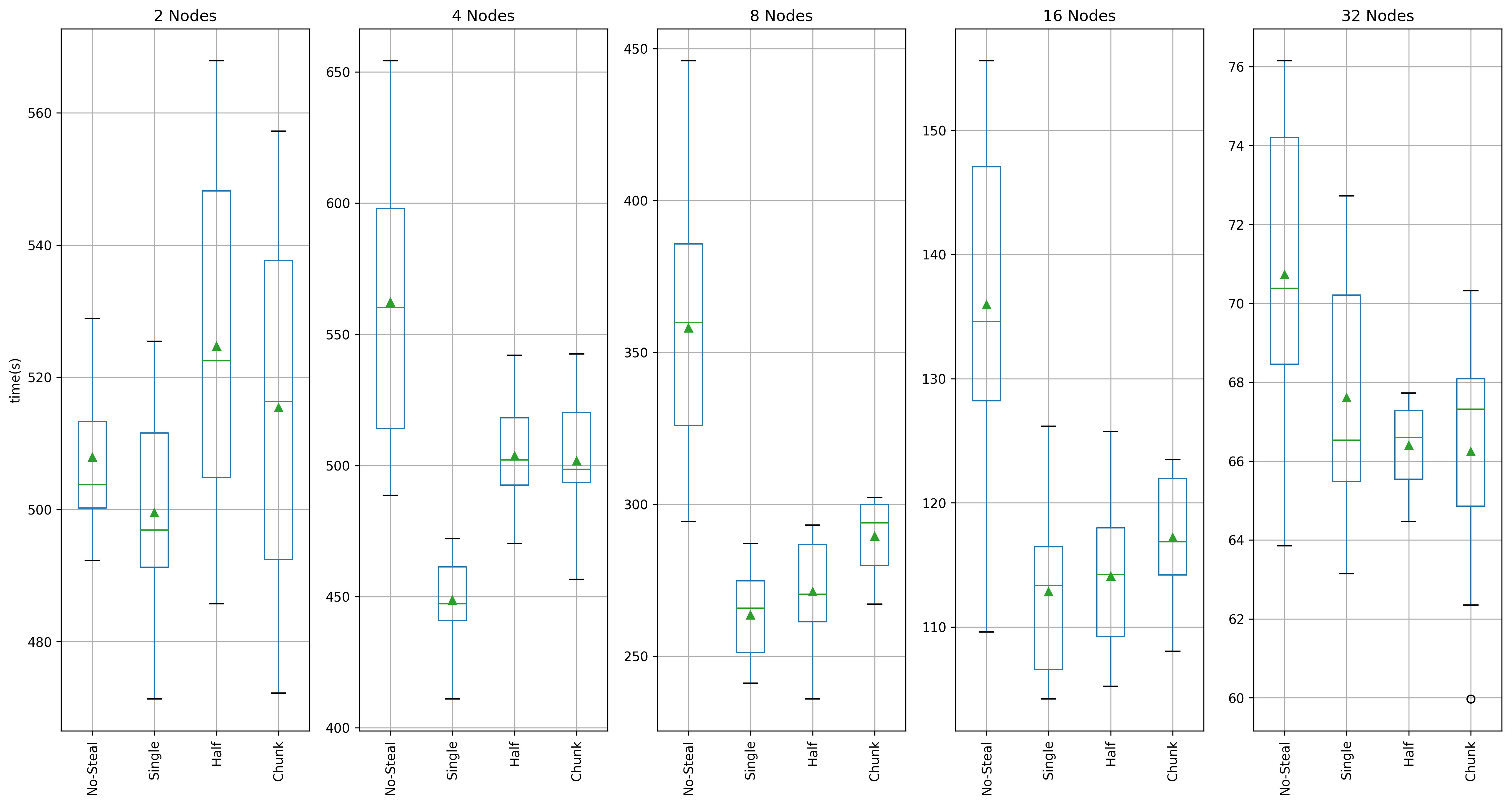}
  \caption{Execution time for different victim policies on varying number of nodes. (Global matrix of $10000^2$ 64-bit elements organized as $200^2$ tiles of $50^2$ elements.)}
  \small
  \label{fig: mig_policy_N}
\end{figure}

The previous experiments showed that work stealing reduces the variation in execution across multiple runs. We postulated that variation occurs because all threads are competing to extract tasks from the scheduling queues. Thus, if the number of threads is large, the queues will be under significant stress, and all the locks will be conflicted leading to large variation in the task acquisition, and thus in the task execution. The scheduler used here use node level queues that are ordered by priority, so the \textit{select} operation can only be done sequentially on all threads. Additionally, in sparse Cholesky factorization, there are a substantial number of tasks that do not do any useful computation, as they are operating on a sparse tile. In such cases, the threads will be spending more time waiting to extract the work, when compared to actual task execution. Fig.~\ref{fig: mig_policy_N} shows the execution time for different victim policies across different numbers of nodes for multiple runs and it shows that work stealing reduces the variation in the execution time. 

\setlength{\belowcaptionskip}{-10pt}

\begin{figure}[htbp]
  \centering
  \includegraphics[width=\textwidth]{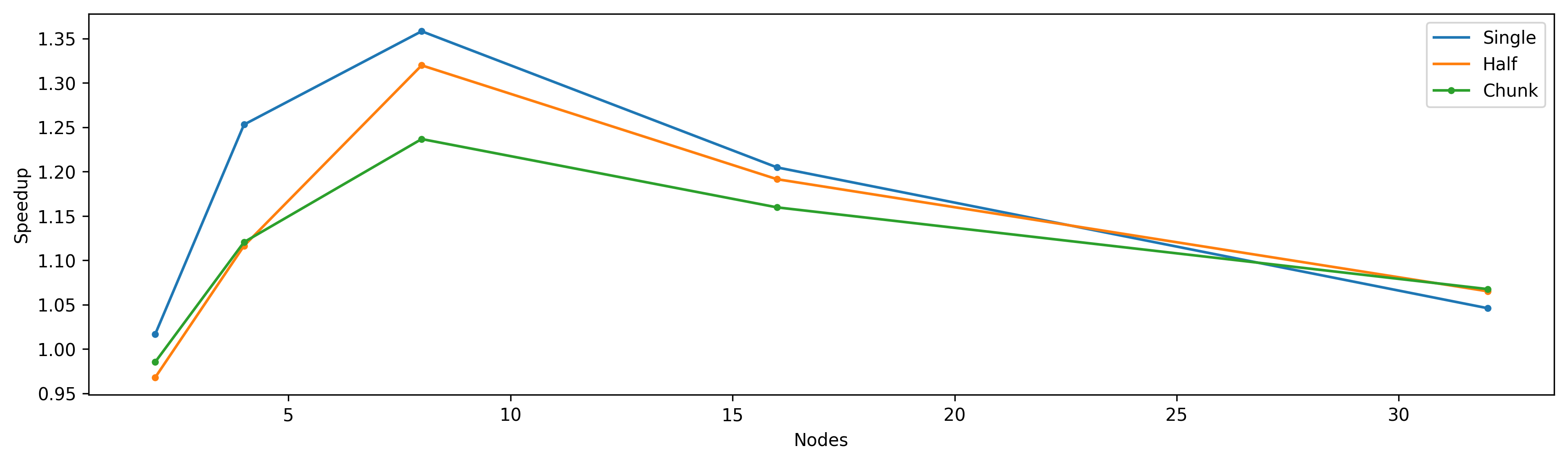}
  \caption{Speedup for different victim policies on varying number of nodes. (Global matrix of $10000^2$ 64-bit elements, organized as $200^2$ tiles of $50^2$ elements)}
  \small
  \label{fig: Speedup}
\end{figure}

The speedup from work stealing (against `No-Steal' as the baseline is not uniform across different numbers of nodes as shown in Fig.~\ref{fig: Speedup}. For each victim policy, speedup is highest (35\%) for 8 nodes, as the potential for work stealing is high (see Fig.~\ref{fig: instant_effective_all}). The speedup decreases for larger number of nodes as the potential for work stealing decreases.
\setlength{\belowcaptionskip}{-10pt}

\subsubsection{Waiting Time}

\begin{figure}[t]
  \centering
  \includegraphics[width=.5\textwidth]{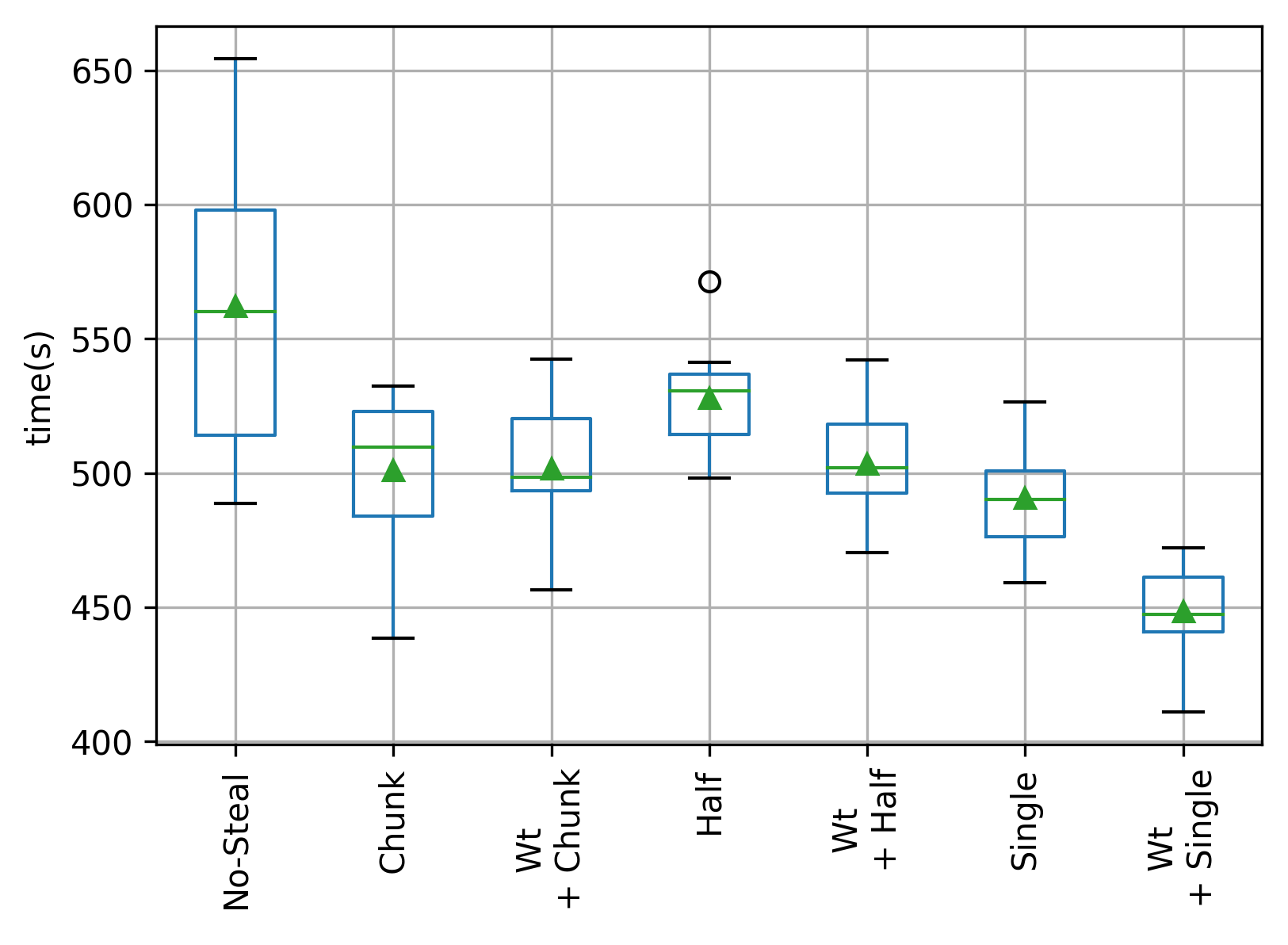}
  \caption{Execution time for different victim policies, with and without waiting time taken into consideration. (Global matrix of $10000^2$ 64-bit elements, organized as $200^2$ tiles of $50^2$ elements)}
  \small
  \label{fig: real_and_wt}
\end{figure}

In all the above experiments, victim policies permit a steal only if the waiting time to execute a task is more than the time taken to steal the task. Fig.~\ref{fig: real_and_wt} shows the comparison in performance when waiting time is taken into consideration and when it is not. Waiting time does not seem to affect \textit{Chunk}, as the mean execution times with and without considering waiting time are similar. Conversely, waiting time has a significant effect on \textit{Half} and \textit{Single}. 

\setlength{\belowcaptionskip}{-10pt}

\begin{figure}[t]
  \centering
  \includegraphics[width=.5\textwidth]{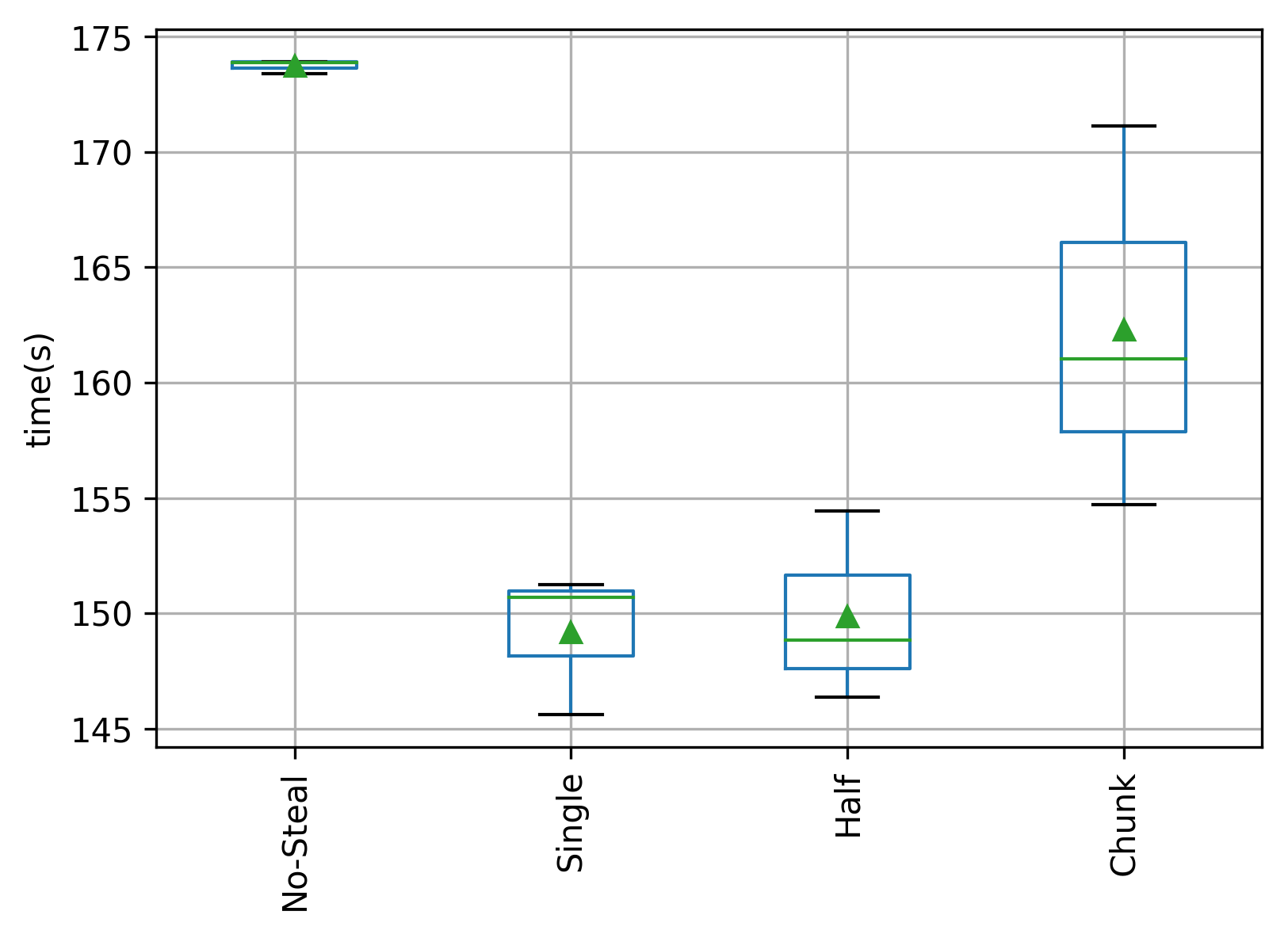}
  \caption{Execution time for different victim policies when using UTS benchmark (b=120, m=5, q= 0.200014, g=$12*10^6$).}
  \small
  \label{fig: uts-exectime}
\end{figure}

In previous work, Perarnau et al.~\cite{Perarnau2014} found that \textit{Half} gives three times the performance of \textit{Chunk} for the Unbalanced Tree Search (UTS) benchmark when waiting time is not considered.  UTS has the property that a child task is always mapped to the same node as its parent task unless stolen by a thief. Due to this mapping property, \textit{Half} makes sense in UTS as no new task will be generated on a starving node. At the same time, there can be an exponential increase in tasks in a busy node. Also, UTS will not suffer from the same problems demonstrated in Fig.~\ref{fig: thief node}, as no new tasks are generated in a starving node. We were able to achieve similar results for UTS (Fig.~\ref{fig: uts-exectime}) but the performance of \textit{Half} was not transferred Cholesky factorization (Fig.~\ref{fig: real_and_wt}). We also found that \textit{Single} has comparable performance to \textit{Half} when using UTS.

\setlength{\belowcaptionskip}{-10pt}

Experiments we conducted using sparse Cholesky factorization (Fig.~\ref{fig: real_and_wt}) show that when waiting time is not considered \textit{Half} performs worse than \textit{Chunk}. When waiting time is taken into consideration, \textit{Half} performs better than \textit{Chunk}, but not by a huge margin. These experiments suggest that when using workloads that have child tasks with multiple parents located on different nodes, it is better to consider waiting time in victim policies. The experiments also demonstrate that if a victim policy gives good performance on one workload, it is not guaranteed that it will deliver similar performance on another.

\subsubsection{Granularity}
Granularity is the time taken to execute a single task. The granularity of different task types may be different but in sparse Cholesky factorization, the granularity of all task types is proportional to the tile size. So we tested the performance of different victim policies against different tile sizes. Table~\ref{table: block_speedup} show that work stealing is more effective with increasing granularity. Also, for smaller granularity, \textit{Chunk} outperforms \textit{Half}. Additionally, for small granularity, work stealing using \textit{Half} actually degrades performance.
\subsubsection{Steal Success Percentage}
\begin{figure}[ht]
  \centering
  \includegraphics[width=\textwidth]{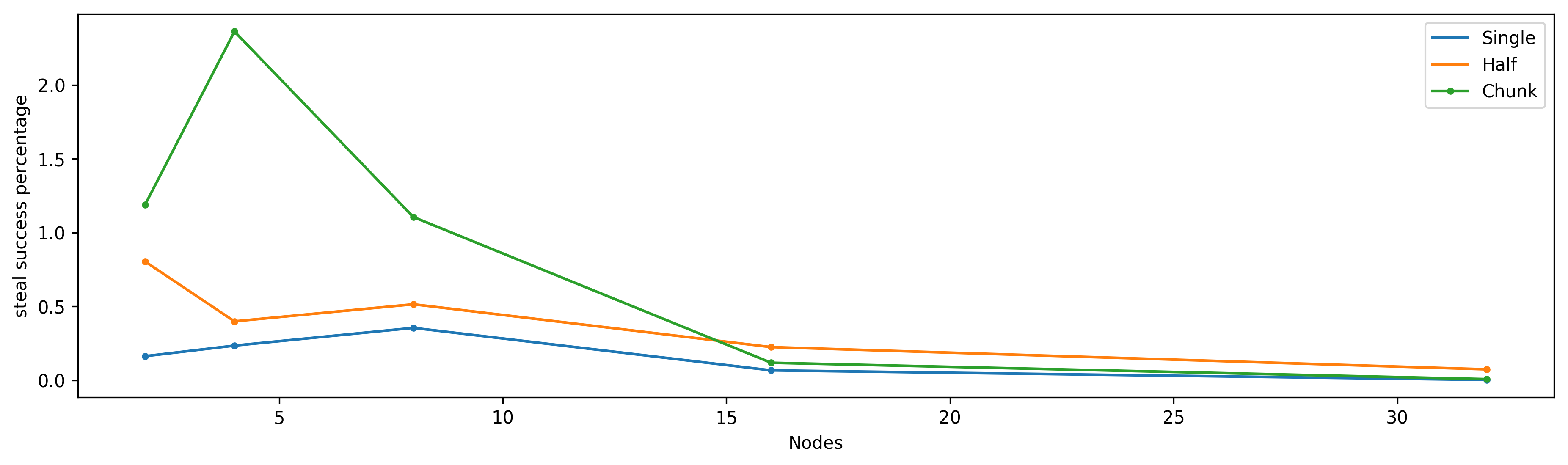}
  \caption{Steal success percentage of different victim policies on varying numbers of nodes. (Global matrix of $10000^2$ 64-bit elements, organized as $200^2$ tiles of $50^2$ elements)}
  \label{fig: steal_success_perc_2}
\end{figure}

\textit{Steal success percentage} is the percentage of steal requests that have yielded at least one task. Fig.~\ref{fig: steal_success_perc_2} shows the steal success percentage for different victim policy. When imbalance is high, steal success is the highest for \textit{Chunk}. At the same time, Fig.~\ref{fig: Speedup} shows that the speedup is highest for \textit{Single} when imbalance is high. From both these experiments, we can conclude that stealing more tasks does not guarantee better speedup, even when there is a high imbalance.
 
\begin{table}[t]
\centering
\caption{Speedup for different victim policies for different tile sizes. (tiled matrix, $10000^2$ tiles, four nodes)}
\label{table: block_speedup}
\begin{tabular}{c|cccc|ccc}
\toprule
& \multicolumn{4}{c|}{\textbf{Execution Time}} & \multicolumn{3}{c}{\textbf{Speedup}} \\
\textbf{Tile size} & \textbf{No-Steal} & \textbf{Chunk} & \textbf{Half} & \textbf{Single} & \textbf{Chunk} & \textbf{Half} & \textbf{Single} \\ 
\midrule

\textbf{10x10}                                                & 230            & 214                                                                  & 244                                                                 & 221                                                                   & 1.077                                                            & 0.94{ }                                                            & 1.03{ }                                                              \\ 
\textbf{20x20}                                                & 237            & 235                                                                  & 246                                                                 & 228                                                                   & 1.006                                                            & 0.96{ }                                                            & 1.03{ }                                                              \\
 
\textbf{30x30}                                                & 255            & 246                                                                  & 253                                                                 & 238                                                                   & 1.03{ }                                                             & 1.008                                                           & 1.07{ }                                                              \\
\textbf{40x40}                                                         & 400            & 370                                                                  & 388                                                                 & 370                                                                   & 1.08{ }                                                             & 1.032                                                           & 1.08{ }                                                              \\
 
\textbf{50x50}                                                & 562            & 501                                                                  & 503                                                                 & 448                                                                   & 1.12{ }                                                             & 1.11{ }                                                            & 1.25{ }                                                              \\
\bottomrule
\end{tabular}
\end{table}

%% file: paper/conclusion.tex
\section{Conclusion}
\label{section: conclusion}

In this paper, we showed that work stealing is an effective load balancing strategy in task-based dataflow runtime, delivering a speedup of up to 35\% and reducing variability in execution time. We also demonstrate that stealing more tasks does not guarantee better speedup, even when there is a high imbalance. \textit{When} the task is stolen is more important than \textit{how many} tasks are stolen and counting future tasks is critical in determining starvation in a thief policy. These experiments suggest that when using workloads that have child tasks with multiple parents located on different nodes, it is better to consider waiting time in victim policies. As an extension of this work, we will be exploring work stealing between accelerator devices in the same node. 

%% file: paper/acknowledgement.tex
\section{Acknowledgement}
This research is undertaken with the assistance of resources and services from the National Computational Infrastructure (NCI), which is supported by the Australian Government. We thank George Bosilca and Thomas Herault (Innovative Computing Laboratory, UTK) for the detailed design discussions.

%% file: bibliography.tex
\bibliographystyle{splncs04}
\bibliography{bibliography}